\documentclass[sigconf]{acmart}
\usepackage{enumitem}
\usepackage{subfigure}

\usepackage{amssymb,amsmath,amsthm}
\usepackage{graphicx}

\usepackage{multirow}
\usepackage{mathtools}
\usepackage{ulem}
\usepackage{hyperref}
\usepackage{comment}
\usepackage{algorithm}
\usepackage{algorithmicx}
\usepackage{algpseudocode}
\usepackage{bm}
\usepackage{bbm}
\newcommand\Set[2]{\{\,#1\mid#2\,\}}

\AtBeginDocument{%
  \providecommand\BibTeX{{%
    \normalfont B\kern-0.5em{\scshape i\kern-0.25em b}\kern-0.8em\TeX}}}

\copyrightyear{2023}
\acmYear{2023}
\setcopyright{acmlicensed}\acmConference[WWW '23]{Proceedings of the
ACM Web Conference 2023}{May 1--5, 2023}{Austin, TX, USA}
\acmBooktitle{Proceedings of the ACM Web Conference 2023 (WWW '23),
May 1--5, 2023, Austin, TX, USA}
\acmPrice{15.00}
\acmDOI{10.1145/3543507.3583363}
\acmISBN{978-1-4503-9416-1/23/04}

\newcommand{\ie}{\emph{i.e., }}
\newcommand{\eg}{\emph{e.g., }}

\newcommand{\etc}{\emph{etc.}}
\newcommand{\wrt}{\emph{w.r.t. }}
\newcommand{\cf}{\emph{cf. }}

\begin{document}

\title{Adap-$\tau$: Adaptively Modulating Embedding Magnitude for Recommendation}

\author{Jiawei Chen}
\authornote{Both authors contributed equally to this research.}
\authornote{Work mostly done at the University of Science and Technology of China (USTC)}
\email{leepyhunt@zju.edu.cn}
\orcid{0000-0002-4752-2629}
\affiliation{%
  \institution{Zhejiang University}
  \country{China}
}

\author{Junkang Wu}
\authornotemark[1]
\email{jkwu0909@gmail.com}
\orcid{0000-0001-6663-926X}
\affiliation{%
  \institution{University of Science and Technology of China}
  \country{China}
}

\author{Jiancan Wu}
\email{wujcan@gmail.com}
\orcid{0000-0002-6941-5218}
\affiliation{%
  \institution{University of Science and Technology of China}
  \country{China}
}

\author{Sheng Zhou}
\email{zhousheng_zju@zju.edu.cn}
\orcid{0000-0003-3645-1041}
\affiliation{%
  \institution{Zhejiang University}
  \country{China}
}

\author{Xuezhi Cao}
\email{caoxuezhi@meituan.com}
\orcid{0000-0002-7044-1341}
\affiliation{%
  \institution{Meituan}
  \country{China}
}

\author{Xiangnan He}
\authornote{Xiangnan He is the corresponding author.}
\email{xiangnanhe@gmail.com}
\affiliation{%
  \institution{University of Science and Technology of China}
  \country{China}
}






\begin{abstract}
   Recent years have witnessed the great successes of embedding-based methods in recommender systems. Despite their decent performance, we argue one potential limitation of these methods --- the embedding magnitude has not been explicitly modulated, which may aggravate popularity bias and training instability, hindering the model from making a good recommendation. It motivates us to leverage the embedding normalization in recommendation. By normalizing user/item embeddings to a specific value, we empirically observe impressive performance gains (9\% on average) on four real-world datasets. Although encouraging, we also reveal a serious limitation when applying normalization in recommendation --- the performance is highly sensitive to the choice of the temperature $\tau$ which controls the scale of the normalized embeddings.
  
    To fully foster the merits of the normalization while circumvent its limitation, this work studied on how to adaptively set the proper $\tau$. Towards this end, we first make a comprehensive analyses of $\tau$ to fully understand its role on recommendation. We then accordingly develop an adaptive fine-grained strategy Adap-$\tau$ for the temperature with satisfying four desirable properties including adaptivity, personalized, efficiency and model-agnostic. Extensive experiments have been conducted to validate the effectiveness of the proposal. The code is available at \url{https://github.com/junkangwu/Adap_tau}. 
\end{abstract}

\ccsdesc[500]{Information systems~Recommender systems}
\ccsdesc[500]{Information systems~Collaborative filtering}
\keywords{Recommendation system, Temperature, Adaptiveness}



\maketitle
\section{Introduction}
Being able to provide personalized suggestions, recommender system (RS) has been widely applied in numerous applications such as social media \cite{covington2016deep,qiu2018deepinf}, advertising \cite{he2014practical,huang2020embedding} and E-commerce \cite{zhou2018deep}. Representation learning is a common paradigm in recommendation, ranging from early matrix factorization (MF) \cite{pan2008one} to recent advanced graph-based models \cite{he2020lightgcn, ying2018graph,wang2019neural}. It learns user/item representation (\ie embeddings) from the historical interactions and then makes a prediction based on the embedding similarity. 
Inner product inherited from MF has been widely applied for measuring embedding similarity, not only because it achieves competitive performance in practice but supports efficient retrieval.

Despite the success, we argue that existing embedding-based methods may not be sufficient for generating satisfactory recommendation --- \ie they do not explicitly modulate the embedding magnitude, which may incur two serious problems, as revealed in our both theoretical and empirical analyses: 1) The free-varying magnitude potentially aggravates the popularity bias. Specifically, we find that the embedding magnitude of popular items grows much more quickly than unpopular items. Those popular items usually exert excessive contribution to model training and finally obtain undesirable higher scores. 2) The highly diverse magnitude hurts model convergence. Through our visual analysis, it is found that even with a proper regularizer, the magnitude of item embeddings is still in a state of rising rather than converging even with numerous epochs.


Being aware of the weaknesses of uncontrolled embedding magnitude, it would be natural to leverage embedding normalization in recommendation. Although embedding normalization has been touched a lot in other fields \cite{wang2017normface, he2020momentum}, it is less explored in RS. By normalizing user/item embeddings into a specific value \wrt temperature $\tau$ \footnote{Instead of directly introducing a parameter for controlling the scale of the normalization, here we refer to recent work that usually utilizes a temperature.} (\ie $1/\sqrt\tau$), we observe impressive performance gains ranging from 5\% to 20\% on four benchmark real-world datasets. Although encouraging, we also reveal a limitation of applying normalization in recommendation --- the performance is highly sensitive to the choice of the temperature $\tau$ that controls the scale of the normalized embeddings. Even a small fluctuation (\eg 0.04) would cause a dramatic performance reduction (sometimes over 10\%). Worse still, the proper $\tau$ may evolve with the data and model shift. Finding the optimal $\tau$ could be extremely hard, which involves a tedious and expensive grid search, heavily hindering the application of the normalization.

To fully foster the merits of the normalization and circumvent its limitation, this work studied an unexplored problem --- \textit{how to adaptively set the proper $\tau$ without requiring notorious hyperparameter tuning}. Towards this end, we first make comprehensive analyses of $\tau$ and reveal its two important roles in model learning:
\begin{itemize}[leftmargin=*]
\item Leveraging temperature could adjust the magnitude of the gradient, while too small or too large $\tau$ would both increase the risk of gradient vanishment.
\item The temperature $\tau$ balances the contributions from the hard negative instances and easy instances. A smaller $\tau$ would make the model pay more attention to hard negative items while a larger $\tau$ make the model treat them equally.
\end{itemize}

Being aware of the role of the temperature in recommendation, we deduce the following two principles to guide the design of the adaptive strategy:


\textbf{Principle 1.} \textit{Adaption principle: temperature should be adaptive to avoid gradient vanishing.}

Principle 1 corresponds to the core role of $\tau$ — avoiding gradient vanishment. Considering the gradient could vary widely with the data distribution and the model changing, $\tau$ should be adapted accordingly.  


\textbf{Principle 2.} \textit{Fine-grained principle: it is beneficial to specify the temperature in a user-wise manner — i.e., the harder the samples of a user are distinguished, the larger the temperature should be employed for the user.}

{Principle 2 is motivated by the hard-mining property. Note that in a typical RS, the data quality usually varies greatly from user to user \cite{eirinaki2013trust}. For the users with much noisy feedback, the model should be more conservative with lifting $\tau$, as the hard samples are likely to be abnormal. Instead, for those users whose feedback is clear and sufficient, lowering $\tau$ to be more aggressive could bring more informative items and enhance model convergence and discrimination. As such, we believe that fine-grained $\tau$ that can adapt to the diverse hardness of different users would be better.}

Based on the aforementioned principles, we propose an adaptive fine-grained strategy named Adap-$\tau$ for specifying the temperature. Towards Principle One, we delve into a global benchmark temperature that maximizes the cumulated magnitude of the gradients. The task is non-trivial, as conventional optimization would involve nested iteration and heavy traversal, incurring serious efficiency problems. Thus, here we develop a skillful approximation and derive a simple close-formed solution for acceleration. 
Towards principle two, we monitor the loss for each user and adapt the temperature accordingly --- a larger loss suggests that samples of the user are hard to be differentiated, which would adaptively increase the value of $\tau$ to reduce the difficulty.  

Finally, we emphasize that Adap-$\tau$ has the following desirable advantages: 1) Adaptability: it is fully adapted to different datasets without requiring any hyper-parameter searching about $\tau$. 2) Personalized: it gives a personalized $\tau$ that can adapt to the diverse hardness of different users. 3) Efficiency: it just involves simple computation without requiring any extra iteration. 4) Model-agnostic: it can be easily plugged in existing embedding-based methods (\eg MF \cite{pan2008one}, LightGCN \cite{he2020lightgcn}) with few codes amended.





\textbf{Contributions.} We summarize the contributions as below:
\begin{itemize}[leftmargin=*]
    \item Revealing the essential of leveraging embedding normalization in recommendation, and identifying one potential limitation --- performance is highly sensitive to the temperature $\tau$. 
    \item Conducting thorough analyses of the temperature $\tau$, uncovering its two important roles in model training and delivering two principles for the temperature specification. 
    \item Proposing an adaptive fine-grained strategy for the temperature with satisfying the four desirable properties including adaptivity, personalized, efficiency, and model-agnostic.
    \item Conducting extensive experiments on four datasets to demonstrate the superiority of our proposal in multiple aspects of recommendation accuracy, adaptivity, and efficiency.
\end{itemize}


\section{Preliminaries}
In this section, we present some background of recommendation.

 
\textbf{Task Formulation.} 
Suppose we have a recommender system with a user set $\mathcal U$ and an item set $\mathcal I$. Let $n$ and $m$ denote the number of users and items in RS. The collected implicit feedback can be expressed by a matrix $Y \in\{0,1\}^{n \times m}$, whose element $y_{ui}$ represents whether a user $u$ has interacted (\eg click) with an item. For convenience, we collect the whole positive instances as $\mathcal D \equiv\{(u,i)|y_{ui}=1\}$; and the positive items (users) for each user $u$ (item $i$) as $\mathcal P_u \equiv\{i|y_{ui}=1\}$ ($\mathcal P_i \equiv\{u|y_{ui}=1\}$). The task of RS is to recommend items for each user that he may be interested in.

\textbf{Embedding-based Model.} Embedding-based methods are widely utilized in RS. They would first transform user/item features (\eg IDs) into vectorized representations (\ie $\mathbf e_u, \mathbf e_i$), and then make predictions based on the embedding similarity. The widely-used similarity functions include inner product \cite{he2016fast} and neural network \cite{he2017neural}. As suggested by recent work \cite{wu2021self,yu2022graph,lin2022improving}, the inner product supports highly efficient retrieval and usually exhibits stronger performance. Thus, for convenience, this work simply takes the representative inner product for analyses, \ie model prediction can be expressed as $\hat y_{ui}=\mathbf e_u^\top \mathbf e_i$.

\textbf{Loss function.} There are multiple choices of loss functions for training a recommendation model including pointwise loss (\eg BCE \cite{he2017neural_a,shan2016deep}, MSE \cite{he2017neural_b,koren2008factorization}), pairwise loss(\eg BPR~\cite{rendle2012bpr}) and Softmax loss \cite{wu2022effectiveness}. Recent work \cite{wu2022effectiveness} finds Softmax loss could mitigate popularity bias, achieves great training stability, and aligns well with the ranking metric. It usually achieves better performance than others and thus attracts a surge of interest in recommendation. In addition, Softmax loss can be considered as an extension of commonly-used BPR loss \cite{rendle2012bpr}. As such, we cast Softmax as the representative loss for analyses, which can be formulated as:
\begin{equation}
    \label{ccl_loss}
    \mathcal{L}=-\frac{1}{|\mathcal{D}|}\sum_{(u,i)\in\mathcal{D}} {\frac{\exp{(\hat y_{ui})}}{\sum_{j\in \mathcal I} \exp{(\hat y_{uj})}}}
    \end{equation}
    In practice, recent paper \cite{wu2022effectiveness} often incorporate a log operation\footnote{$\mathcal{L}=-\frac{1}{|\mathcal{D}|}\sum_{(u,i)\in\mathcal{D}} \log {\frac{\exp{(\hat y_{ui})}}{\sum_{j\in \mathcal I} \exp{(\hat y_{uj})}}}$} and conduct negative sampling or mini-batch strategy \cite{oord2018representation} for acceleration. But they are not our focus and here we simply refer to the original loss \cite{covington2016deep} for theoretical analyses.
    \begin{figure}[t]\centering
        \vspace{-.5em}
        \includegraphics[width=1.0\linewidth]{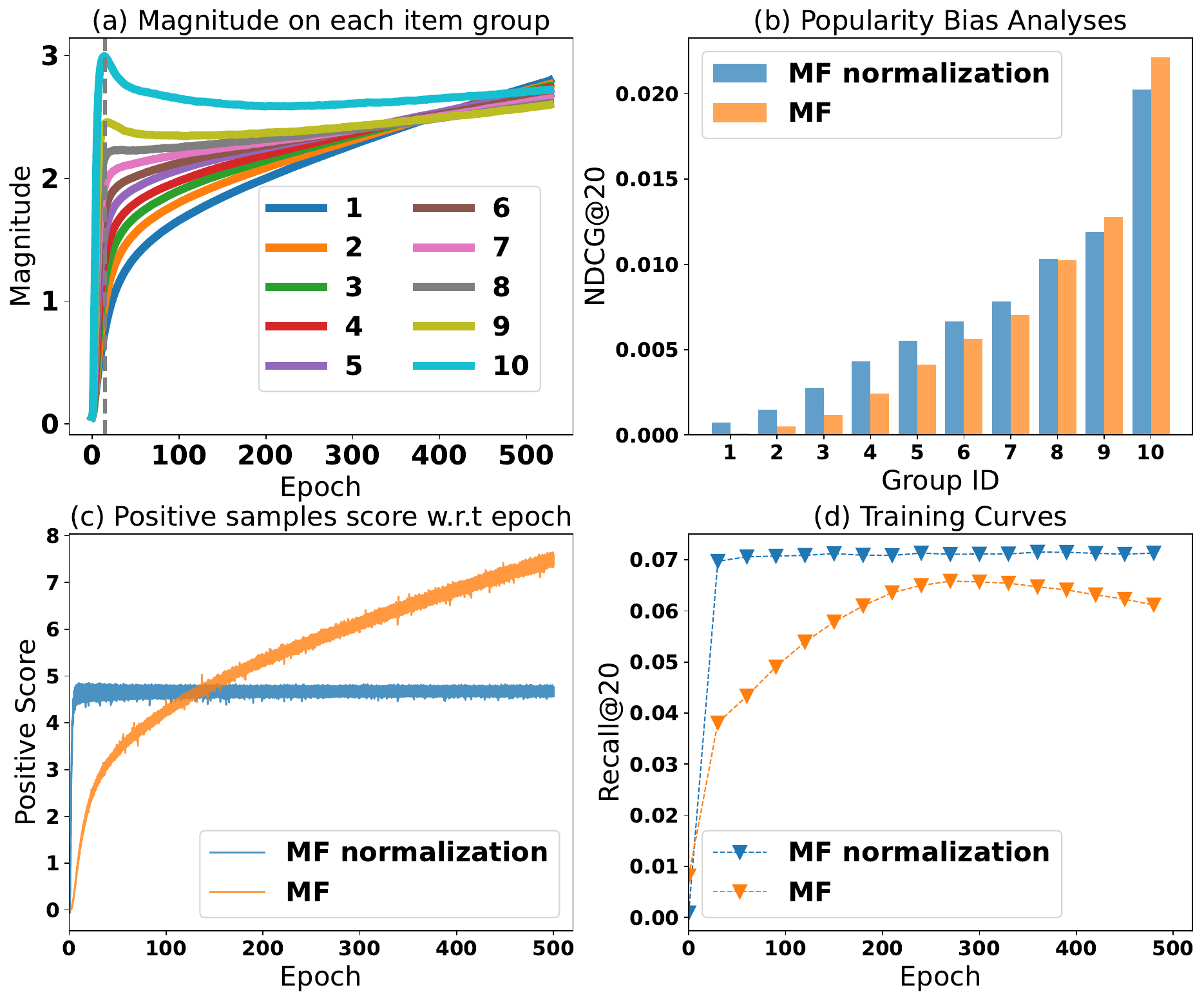}
        \caption{Empirical studies on Yelp2018: Fig. (a) and Fig. (b) represent item embedding magnitude of the different groups across the training procedure and respective performance. The larger GroupID is, the more popular items the group contains. Fig. (c) and Fig. (d) depict the positive samples score and corresponding performance in the training procedure. }
        \label{fig::FIST_FIG}
        \vspace{-.8em}
        \end{figure}
    
    \textbf{Embedding Normalization.} This work studies the nature of embedding normalization in recommendation. On the basis of inner product, we leverage embedding normalization in prediction as:
    \begin{equation}
    \hat y_{ui}=\frac{{{\bf{e}}_u^ \top {{\bf{e}}_i}}}{{\left\| {{{\bf{e}}_u}} \right\|\left\| {{{\bf{e}}_i}} \right\|}} \cdot \frac{1}{\tau }
    \end{equation}
    where the magnitude of user/item embeddings has been rescaled. The first factor:
    \begin{equation}
    f(u,i) =\frac{{{\bf{e}}_u^ \top {{\bf{e}}_i}}}{{\left\| {{{\bf{e}}_u}} \right\|\left\| {{{\bf{e}}_i}} \right\|}}
    \end{equation}
    can be understood as cosine similarity, where the magnitude has been isolated; and the second factor $1/\tau$ rescales the normalized embeddings. We remark that instead of directly introducing a parameter controlling the scale, we borrow a similar idea in contrastive learning \cite{chen2020simple, he2020momentum} and utilize the conventional temperature. The alignment could make our findings better generalized to other domains. 


\section{Analyses over Embedding Normalization}
In this section, we first validate the essential of leveraging embedding normalization in RS (Sec. 3.1), and then identify one potential limitation (Sec. 3.2). Finally, we conduct thorough analyses of the temperature and uncover its two important roles (Sec. 3.3).
\subsection{Necessity of Normalization}
\subsubsection{\textbf{Theoretical Analysis}}
We start with theoretical analysis to show that without normalization the magnitude of popular items grows much more quickly than unpopular items. In fact, we have:
\newtheorem{lem}{Lemma}
\begin{lem}
\label{lam0}
By choosing inner product without controling magnitude, we have change of item embedding magnitude $\delta_{i}$ in each iteration:

\begin{equation} \label{gd_2_2}
    \delta_{i} = \sum_{u \in \mathcal{P}_i} \frac{2}{\tau} (\mathbb{I}[y_{ui}=1] - \sum\limits_{k \in {N_u}} {{p_{uk}(\tau)}} ) p_{ui} f(u,i)
\end{equation}

At the early stage of the training procedure, $\delta_{i}$ obeys:
\begin{equation} \label{gd_2_3}
    \delta_{i} \propto |\mathcal{P}_i|
\end{equation}

where $ p_{ui}(\tau)=({exp(\frac{f(u,i)}{\tau})})/({\sum\limits_{j\in \mathcal I}{\exp(\frac{f(u,j)}{\tau})}})$ and $|\mathcal{P}_i|$ represents the frequency of item $i$, and $\mathcal{P}_i$ denotes the set of users observed in $\mathcal{D}$ which are interactived with $i$.
\end{lem}
The proof of the lemma is presented in Appendix \ref{proof_lama}. We can draw an observation from Lamma \ref{lam0}:
Note that at the early stage of the training procedure, users and items are distributed uniformly. In other words, $p_{ui} f(u,i)$ cannot tell remarkable difference and $\sum\limits_{k \in {N_u}} {{p_{uk}(\tau)}}$ is relatively small, while the magnitude of popular items will obtain explosive rising in term of $|\mathcal{P}_i|$ ($\mathbb{I}[y_{ui}=1]$). 


\begin{figure}[t]\centering
    \vspace{-.5em}
    \includegraphics[width=1.0\linewidth]{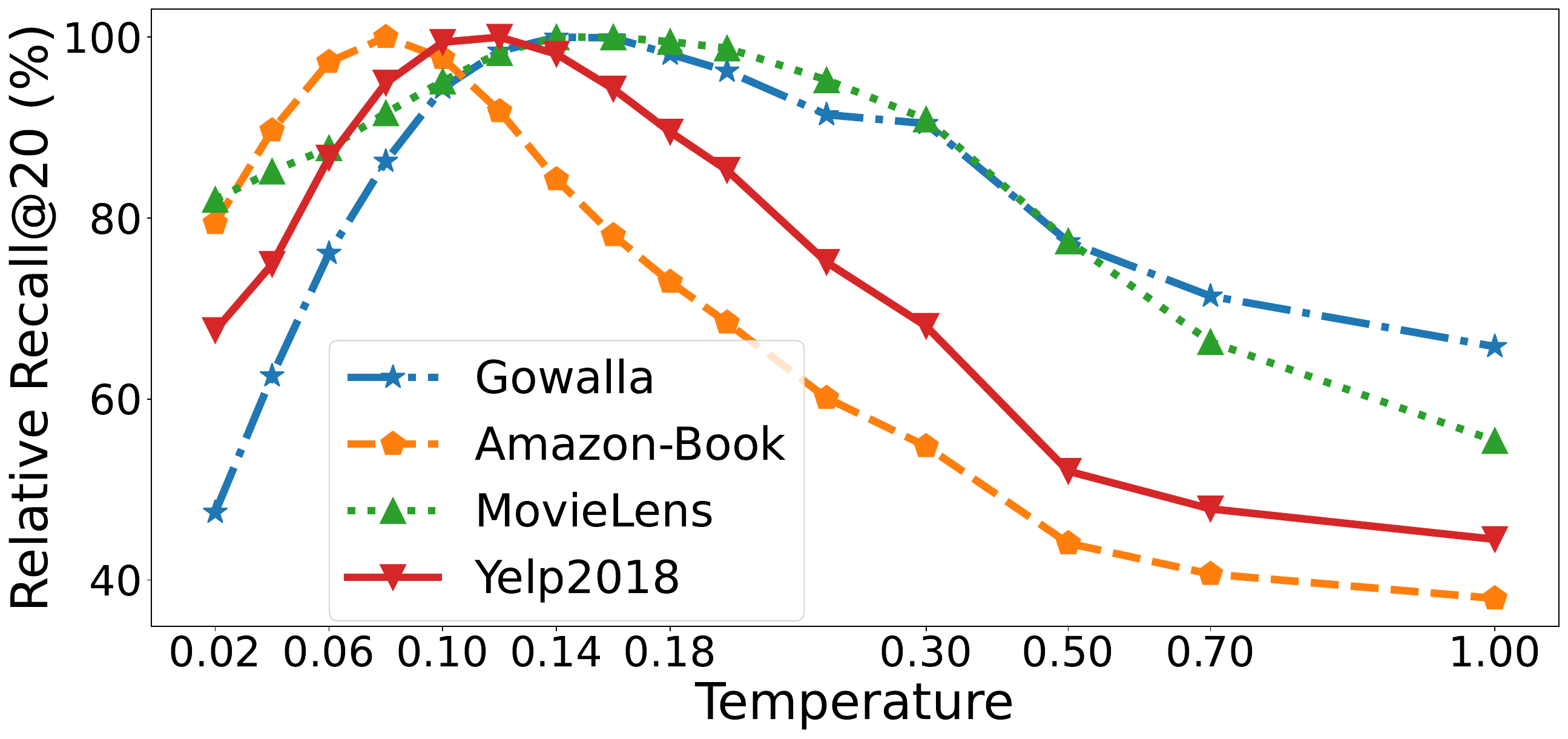}
    \caption{Relative recall@20 over four datasets with $\tau$ .}
    \label{fig::tau_change}
    \vspace{-.8em}
    \end{figure}

    \begin{table}[t]
        \caption{Performance comparisons of MF with/without embedding normalization. The column of "norm" represents whether to conduct normalization for the user or item representation. For example, Y-N stands for adopting normalization on the user side but not on the item side. }
        \label{tab:norm_method}
        \centering
        \begin{tabular}{ccccc}
            \toprule
            \multirow{2}{*}{norm?} & \multicolumn{2}{c}{Yelp2018} &  \multicolumn{2}{c}{Amazon-book} \\
             & Recall  & NDCG  & Recall  & NDCG  \\
            \hline
            N-N & 0.0677 & 0.0554 & 0.0457 & 0.0352 \\
            Y-N & 0.0709  & 0.0585 & 0.0529 & 0.0419 \\
            N-Y & 0.0703  & 0.0577 & 0.0513 & 0.0399  \\
            Y-Y & \textbf{0.0714} & \textbf{0.0586} & \textbf{0.0542} & \textbf{0.0422} \\
            \bottomrule
        \end{tabular}
    \end{table}

    \subsubsection{\textbf{Empirical Analysis}}
    From Lemma \ref{lam0}, we know that the magnitude is correlated with item popularity. In this subsection, we explore its negative impact on recommendation through rich experiments.
    
    \textbf{Experiments design.} To show the impact of free-varying magnitude, here we conduct four experiments: \textbf{(1)} We first visualize the magnitude of item embedding with different item popularity during the training  (Fig. \ref{fig::FIST_FIG} a). Here we follow \cite{wu2022effectiveness} and split items into ten groups in terms of item popularity. The larger group ID indicates the group contains more popular items. \textbf{(2)} We also report the performance in terms of different item groups (Fig. \ref{fig::FIST_FIG} b). \textbf{(3)} The predictive scores of positive instances with training epochs is presented in Fig. \ref{fig::FIST_FIG} c. \textbf{(4)} We visualize the performance of MF with or without normalization (Fig. \ref{fig::FIST_FIG} d). All experiments are conducted on the MF backbone and Yelp2018 \cite{he2020lightgcn} dataset. Similar results can be observed on other models (like LightGCN) and datasets. The details of experimental settings can refer to section \ref{set_exp}. 

    
    
    \textbf{Free-varying magnitude aggravates popularity bias.} If we focus on the early stage of the training (\cf Fig. \ref{fig::FIST_FIG} (a)), the magnitude of popular items rises rapidly which is consistent with theoretical proofs. Therefore, popular items are prone to obtain higher scores as the magnitude directly contributes to model prediction. Besides, the diverse magnitude also hurt the training of the user embedding. The gradient of user embedding can be written as: $\frac{{\partial L}}{{\partial {\mathbf e_u}}} = \sum_{u,i} {\frac{{\partial L}}{{\partial f(u,i)}}{\mathbf e_i}}$, where popular items with larger magnitude would exert excessive
    contribution and potentially overwhelm the signals from others. The model would sink into biased results. Fig. \ref{fig::FIST_FIG} (b) provides the evidence. The model would sink into biased results. As can be seen, the model with normalization yields much fairer results than the model without normalization.

    
    \textbf{Free-varying magnitude hurts convergence.} If we turn our attention to the end of training in Fig. \ref{fig::FIST_FIG} (c), we observe that even with numerous epochs (\eg 500), the predicted scores and embedding magnitude of vanilla MF are still in a state of rising rather than convergence while the performance drop consistently (Fig. \ref{fig::FIST_FIG} (d)). 
    But when we leverage normalization in MF, we observe impressive improvement --- the model arrives at convergence quickly with much fewer epochs (\ie 20) and performs stable with more epochs.  
    
    
    \textbf{Normalization boosts performance.}. To further validate the merit of the normalization, here we directly test the recommendation performance with or without conducting normalization on the user or item embeddings (Table \ref{tab:norm_method}). As can be seen, the model with both-side normalization (\ie Y-Y) remarkably outperforms the model with one-side normalization (\ie Y-N or N-Y); and they both surpass the model without normalization (N-N).

\subsection{Limitation of Normalization}
Although we have proved the superiority of normalization in recommendation tasks, here reveal one potential limitation of the normalization --- the performance is highly sensitive to the temperature $\tau$. 

To validate this point, we test the recommendation performance \wrt $\tau$ ranging from 0.02 to 1 with a rather fine-grained step-size 0.02. The result is shown in Fig. \ref{fig::tau_change}, where we report the relative performance with the best for better visualization. We make the following observations: 1) the performance is highly sensitive to $\tau$. Even a small fluctuation (\eg changing from 0.08 to 0.12 on Amazon-Book) would cause a dramatic performance reduction (\eg 10\%); 2) Different datasets require rather different $\tau$. For example, Amazon-Book dataset reaches the best performance when $\tau=0.08$, but MoiveLens reaches with $\tau=0.16$. If we simply transfer the optimum $\tau$ in one dataset (\eg Moivelens) to another (\eg Amazon-Book), we would get rather poor performance (\eg over 30\% reduction). 

As a result, finding optimal $\tau$ is highly difficult, which heavily hinders the application of embedding normalization. Methods like grid search or automated hyperparameter search \cite{feurer2019hyperparameter} are potential to find the optimum, but they are highly time-cost expensive. As such, we believe it is essential to pursue a automatic mechanism that could specify the proper $\tau$ adaptively.


\subsection{Roles of Temperature}
In this subsection, we make a comprehensive analysis of $\tau$ and reveal its two important roles in model learning. 
\subsubsection{\textbf{Avoiding gradient vanishment.}} The temporature mainly affect the gradient of the loss function $L$ \wrt $f(u,i)$. For convenient, let notation $p_{ui}(\tau)$ be the logit of the instance $(u,i)$ governed by the parameter $\tau$, \ie
\begin{align}
    p_{ui}(\tau)=\frac{exp(\frac{f(u,i)}{\tau})}{\sum\limits_{j\in \mathcal I}{\exp(\frac{f(u,j)}{\tau})}}
\end{align}
The gradient  $\frac{{\partial L}}{{\partial f(u,i)}}$ can be written as:
\begin{align}
    \label{gd11}
        \frac{{\partial L}}{{\partial f(u,i)}} = \left\{ {\begin{array}{*{20}{c}}
           \frac{1}{\tau} {{p_{ui}(\tau)}(1 - \sum\limits_{k \in {\mathcal P_u}} {{p_{uk}(\tau)}} )},  \quad \text{for} \quad y_{ui}=1\\
           - \frac{1}{\tau} {{p_{ui}(\tau)}(\sum\limits_{k \in {\mathcal P_u}} {{p_{uk}(\tau)}} )}, \quad \text{for} \quad y_{ui}=0
            \end{array}} \right.
    \end{align}
The expected magnitude of the graidents can be written as:
\begin{equation}
    \begin{aligned}
    \mathbb{E}_{i}[|\frac{{\partial L}}{{\partial f(u,i)}}|] 
        &=\frac{2}{m\tau}\sum\limits_{i\in  \mathcal P_u}{p_{ui}(\tau)(1-\sum\limits_{k\in \mathcal P_u}{p_{uk}(\tau)})} 
    \end{aligned}
    \label{eq:vanish222}
\end{equation}
which can be understood as the product of the sum of positive logits ($\sum_{i\in  \mathcal P_u}{p_{ui}(\tau)}$) and the sum of the negative logits (1-$\sum_{k\in  \mathcal P_u}{p_{uk}(\tau)}$). When $\tau$ is too small, due to the explosion nature of exponential function, the disparity on $f(u,i)$ would be amplified, and positive instances usually obtain extremely larger logits than negative (\eg $\sum_{i\in  \mathcal P_u}{p_{ui}(\tau) \to 1}$). The gradient would vanish. On the contrary, when $\tau$ is too large, the logits $p_{ui}$ do not exhibit much difference. But due to the long tail nature of RS --- \ie the number of negative instances is much larger than positive, the sum of positive logits would be quite small and the gradient vanishes again. 

Appendix \ref{sec_vannise_proof} provides an example of how the gradient magnitude varies with the temperature $\tau$. As can be seen, too large or too small $\tau$ would cause gradient vanishment. As such, $\tau$ should be specified carefully and adapted for obviating gradient vanishment.  

\subsubsection{\textbf{Hard-mining}}

Hard-mining of $\tau$ has been uncovered by some recent work in contrastive learning \cite{wang2021understanding}. Here we borrow their ideas but provide more insightful analyses in terms of RS scenarios. As discussed before, the exponential function with small $\tau$ would amplify the disparity. Hence those hard negative samples with larger $f(u,i)$ would have extremely larger logits $p_{ui}$, contributing more on model training. On the contrary, larger $\tau$ tends to make the model treat the negative samples equally. 

This property highly motivates us to give a user-wise $\tau$. Note that in a typical RS, the data quality usually vary greatly from user to user. For the users with much noisy feedback, it would be unwise to concentrate much on the hard negative samples, as they are likely to be noisy samples. But for those users with clear and sufficient feedback, lowering $\tau$ would be a better choice as it could bring more informative samples and thus enhances model convergence and discrimination. As such, continuing on the habit of fixed $\tau$ is no longer a wise choice.  It would be better to give fine-grained $\tau$ that can adapt to the diverse hardness of different users.

More interestingly, this treatment bring another advantage. From Eq.(\ref{gd11}), we find the gradient is discounted by $1/\tau$. Giving personalized $\tau$ could also adjust the confidence of users -- \ie the users with higher-quality data would make more contributions on training.

\section{Proposed Method}
To address this problem, in this section, we propose \textit{Adap-$\tau$} that is able to adaptively and automatically modulate the embedding magnitude in recommender system. Ada-$\tau$ is developed based on the following principles:
\begin{itemize}[leftmargin=*]
    \item \textit{(P1) Adaption principle: temperature should be adaptive to avoid gradient vanishing.}
    \item \textit{(P2) Fine-grained principle: it is beneficial to specify the temperature in a user-wise manner --- \ie the harder the samples of a user are distinguished, the larger temperature should be employed for the user.}   
\end{itemize}

\subsection{Adap-$\tau_0$: Towards Adaptive Temperature}
Towards principle (P1), we delve into an automatic temperature that maximizes the magnitude of the gradients:
\begin{equation}
    \tau_{0}=\arg\max_{\tau}\mathbb{E}_{u\in U,i \in I}[|\frac{{\partial L}}{{\partial f(u,i)}}|] 
\label{eq:t0}
\end{equation}
Directly optimizing the equation (\ref{eq:t0}) is computationally infeasible, as it would involve nested optimization and heavy traversal over each user-item pair. Thus, we turn to pursue an efficient approximated solution. Here we first derive the tight upper bound of the objective, which can be easily optimized. In fact, we have:

\begin{lem}
\label{la1}
Let $p_{ui}(\tau)$ be the logit of the instance $(u,i)$ governed by the parameter $\tau$, \ie $p_{ui}(\tau)=exp(f(u,i)/\tau)/\sum_i{\exp({f(u,i)/\tau})}$ and $\tau$ be lower bounded \footnote{In practical, we usually control the lower bound of $\tau$ to guarantee numerical stability.} by T. The objective is bounded with:
\begin{align}
    \mathbb{E}_{u,i}[|\frac{{\partial L}}{{\partial f(u,i)}}|] 
        \leq \frac{2}{mT}(\mathbb E_u[\sum\limits_{i\in N_u}{p_{ui}(\tau)}]-\mathbb E^2_u[\sum\limits_{i\in  N_u}{p_{ui}(\tau)}])
\end{align}
The optimum of the upper bound is achieved iff the following condition holds: 
\begin{align}
    E_u[\sum\limits_{i\in N_u}{p_{ui}(\tau)}]=\frac{1}{2} \label{eq:con}
\end{align}
\end{lem}

The proof of the lemma is presented in Appendix \ref{proof_lemma2}. Now the question lies on solving the Equation (\ref{eq:con}), which is still intractable. Here we explore a reasonable approximate and have:

\begin{lem}
    \label{la2}
    Let $\mathbb F$ (or $\mathbb F^+$) be the distribution of $f(u,i)$ over all instances (or positive instances). Let $\mathbf f$ (or $\mathbf f^+$) be a random variable that sampled from $\mathbb F$ (or $\mathbb F^+$). Suppose the distribution $\mathbb F$ and  $\mathbb F^+$ have a sub-exponential tail such that the following conditions hold for some $\lambda,\lambda_+>0$:
    \begin{equation}
        \begin{aligned}
    &p((\mathbf f-\mathbb E_\mathbb F[\mathbf f]) > b) \le 2{e^{ - 2b/\lambda }} \\
    &p((\mathbf f_+-\mathbb E_{\mathbb F_+}[\mathbf f_+]) > b) \le 2{e^{ - 2b/\lambda_+ }}
\end{aligned}
\end{equation}
When $\tau_0\geqslant \max(2\lambda,2\lambda_+, T)$, it can be approximated as: 
\begin{equation}
    \begin{aligned}
    \label{eq:sol1}
\tau_0\approx \frac{{\sigma _ + ^2 - {\sigma ^2}}}{{ - ({\mu ^ + } - \mu ) + \sqrt {{{({\mu ^ + } - \mu )}^2} + 2(\sigma _ + ^2 - {\sigma ^2})\log (\frac{{nm}}{{2|D|}})} }}
\end{aligned}
\end{equation} 
    where $|D|$ denotes the number of positive instances in the datasets, $\mu$ (or $\mu_+$) and $\sigma^2$ (or $\sigma^2_+$) denotes the mean and variance of $\mathbf f$ (or $\mathbf f_+$). 
when $\sigma^2 _ +$ is close to $\sigma^2$ (\cf Appendix \ref{proof_apendxi}), the expression can be simplified as:
\begin{equation}
    \begin{aligned}
        \tau_0\approx\frac{\mu_{+}-\mu}{\log(\frac{nm}{2|D|})} \label{eq:sol}
    \end{aligned}
\end{equation}

\end{lem}
The proof of the lemma is presented in Appendix \ref{proof_lemma3}. 
Here we make a hypothesis on the distribution --- \ie $\mathbb F$ and $\mathbb F_+$ are convergent and the tails of the distribution decay at least as fast as exponential one (that decay as $e^{-2t/\lambda}$). The hypothesis is practical as the sub-exponential random variables is actually common. It contains Guassian, exponential, Gamma, Pareto, Cauchy, \etc. Besides, Hoeffding \cite{block1988multivariate} proofed that all bounded random variables are sub-exponential. 

In fact, in our experiments, we always observe that $f$ and $f_+$ are convergence into a specific region, with a pretty small $\lambda$ and $\lambda_+$. Also, we observe that the two distribution usually has a quite close variance (\cf Appendix \ref{proof_apendxi}). These observations validate the Equation (\ref{eq:sol}) can be safely applied. Our empirical study presented in Section \ref{exp} also validate the superiority of the proposed strategy.

\begin{table*}[t]
    \caption{Performance comparison between Adap-$\tau$ and other similar strategy. `No-Norm' denoted the method without normalization. `Grid Search $\tau$' denoted the method leveraging grid search to find optimal $\tau$. `C-$\tau$' and `Cu-$\tau$' utilize the neural network to learn $\tau$ following the work \cite{wang2017normface}.}
    \label{tab:main}
    \centering
    \begin{tabular}{cccccccccc}
        \toprule
        \multirow{2}{*}{Backbone} & \multirow{2}{*}{strategy} & \multicolumn{2}{c}{Yelp2018} & \multicolumn{2}{c}{Amazon-book} &
        \multicolumn{2}{c}{Movielens} & \multicolumn{2}{c}{Gowalla}\\
        \cline{3-10} 
        \multicolumn{2}{c}{} & Recall  & NDCG  & Recall  & NDCG & Recall  & NDCG & Recall  & NDCG \\
        \hline
        \multirow{6}{*}{MF} & No Norm & 0.0677 & 0.0554 & 0.0457 & 0.0352 & 0.2721  & 0.2525 &  0.1616 & 0.1366   \\
        & Grid Search $\tau$ & 0.0714  & 0.0586 & 0.0542 & 0.0422  & 0.2789 & 0.2624  & 0.1761 & 0.1399\\
        \cline{2-10}
        & C-$\tau$ & 0.0647  & 0.0528 & 0.0538 & 0.0418 & 0.2472 & 0.2260 &  0.1723&  0.1362 \\
        & Cu-$\tau$ & 0.0691  & 0.0566   & 0.0541 & 0.0421  & 0.2600 & 0.2398  &  0.1751 & 0.1383 \\
        \cline{2-10}
        & Adap-$\tau_0$ & 0.0714  & 0.0585 & 0.0549 & 0.0427  & 0.2792 & 0.2638 & 0.1754 & 0.1386  \\
        & Adap-$\tau$ & \textbf{0.0721}  & \textbf{0.0594}  & \textbf{0.0553}& \textbf{0.0430} & \textbf{0.2815} & \textbf{0.2673} & \textbf{0.1838} & \textbf{0.1506} \\
        \hline
        \multirow{6}{*}{LightGCN} & No Norm & 0.0649 & 0.0530  &  0.0411 & 0.0315  & 0.2576 & 0.2427 &  0.1830 & 0.1554  \\
        & Grid Search $\tau$ & 0.0730  & 0.0605 & 0.0596 & 0.0477 & 0.2767  & 0.2575 & 0.1878  & 0.1577 \\
        \cline{2-10}
        & C-$\tau$ &  0.0653 & 0.0537  & 0.0571 & 0.0453 & 0.2529 & 0.2282 & 0.1731 & 0.1431  \\
        & Cu-$\tau$ &  0.0690 & 0.0571 & 0.0586 & 0.0468 & 0.2582 & 0.2357 & 0.1797 & 0.1488 \\
        \cline{2-10}
        & Adap-$\tau_0$ & 0.0724  & 0.0603 & 0.0601 & 0.0480 & 0.2744 & 0.2571 & {0.1841} & 0.1526 \\
        & Adap-$\tau$ & \textbf{0.0733}  & \textbf{0.0612} & \textbf{0.0612} & \textbf{0.0490} & \textbf{0.2787} & \textbf{0.2615} &\textbf{0.1901} & \textbf{0.1590} \\
        \bottomrule
    \end{tabular}
\end{table*}

\subsection{Adap-$\tau$: Towards Adaptive Fine-grained Temperature}
Towards principle (2), we introduce personalized temperatures $\tau_u$ for each user and leverage a \textit{Superloss} \cite{NEURIPS2020_2cfa8f9e} to supervise their learning. Specifically, the role of Superloss is to monitor the loss of the samples for each user, and to adaptively adjust the temperature values accordingly. It is composed of a loss-aware term and a regularization term: 
\begin{equation}
    \begin{aligned} 
J = \frac{L(u)-m_u}{\tau _u} + \beta {(\log {\tau _u} - \log {\tau _0})^2} \label{eq:sl}
\end{aligned}
\end{equation}
where $L(u)$ denotes the cumulated loss of the samples of a specific $u$, reflecting how difficult the samples to be differentiated by the model. $m_u$ is a threshold that ideally separates easy samples from hard samples based on their respective loss, which can be set as the mean of the $L(u)$ in practical. The larger $L(u)$ would bring the more penalty, which would reduce the $1/\tau_u$ to a larger extent and adaptively pushes the temperature towards larger value. Remarkably, to make a fair comparison of $L(u)$ over users, here we choose the common temperature to calculate $L(u)$.  

Meanwhile, to prevent the temperature from sinking into extreme values that incurs gradient vanishing, a regularizer has also been introduced. This regularizer aims at pulling the learned $\tau_u$ to be close to the $\tau_0$ that has approximately largest gradient magnitude. A parameter $\beta$ is introduced to balance both effects. It can be simply set as  $1.0$ without requiring grid search. Instead of learning the optimal $\tau_u$ via back-propagation, we prefer to find the close-formed solution, which does not involve extra iteration causing efficiency issue or notorious fine-tuning on the extra parameters of the learning rate or decay. In fact, we have a closed-form solution from Eq. ($\ref{eq:sl}$):
\begin{equation}
    \begin{aligned} 
\tau _u^* = \tau_0 \cdot \exp (\mathbb W(\max ( - \frac{1}{e},\frac{{{L(u)} - {m_u}}}{{2\beta }}))) \label{eq:lw}
\end{aligned}
\end{equation}
where $\mathbb W(.)$ stands for the Lambert-W function, which is an inverse function of $xexp(x)$. As intended, $\tau_u^*$ is monotonically increasing with the user loss $L(u)$ --- the user with a larger loss would acquire a larger $\tau_u$ to down weight the confidence of the user. Meanwhile, the $\tau_0$ acts as a baseline to scale the $\tau$ into a proper region.   

\subsection{Discussion}
We show that the proposed Adap-$\tau$ satisfies the following three desirable properties:

\textbf{Personalization.} As for user-wise adaption, owing to the capability of \textit{Superloss}, our model could calculate the specific $\tau$ in terms of their cumulated loss. The larger train loss suggests the data may contain more noises, and thus drives the model to be more conservative. $\tau$ would become larger accordingly to slow the pace of hard-mining and down-weigh the contribution of this user.

\textbf{Adaption.} As for data-wise adaption, our Equation (\ref{eq:t0}) automatically computes proper $\tau$ without extra manual intervention, thus avoiding the notorious hyper-parameter search for $\tau$.

\textbf{Model-agnostic:} Actually, our Adap-$\tau$ can be easily plugged in many embedding-based methods. We do not deeply intervene on the model, but simply introduce embedding normalization and adaptive temperatures calculated from Eq. \eqref{eq:lw}.

\textbf{Efficiency:} 
Our calculation about adaptive $\tau$ is a straightforward close-formed without requiring extra iteration. Also, the temperature can be calculated efficiently. As for $\mu^+$, we compute it with element-wise multiplication of positive instances which costs $O(2|\mathcal{D}|d)$, where $|\mathcal{D}||$ denotes the number of positive instances and $d$ represents dimension. And $\mu$ is calculated by the cosine similarity between users and mean value of all items representation for simplicity, which cost $O(nd)$, where $n$ denotes the number of users. Its total complexity is  $O(2|E|d + nd)$ without backward cost. More detailed analyses could refer to Appendix \ref{algorithm_pse}.

\section{Experiments}
\label{exp}
In this section, we present comprehensive experiments to demonstrate the effectiveness of our model. Our experiments are intended to address the following research questions::
\begin{itemize}[leftmargin=*]
    \item \textbf{RQ1:} How does Adap-$\tau$ perform compared with other strategies?
    \item \textbf{RQ2:} Does our Adap-$\tau$ adapt to different datasets and users? 
    \item \textbf{RQ3:} How does the model equipped with embedding normalization and adaptive $\tau$ perform compared with state-of-the-art in terms of both accuracy and efficiency?
\end{itemize}
\subsection{Experimental Settings}
\label{set_exp}
\subsubsection{\textbf{Datasets and Metrics}}
\begin{figure*}[t]\centering
    \vspace{-.5em}
    \includegraphics[width=.8\linewidth]{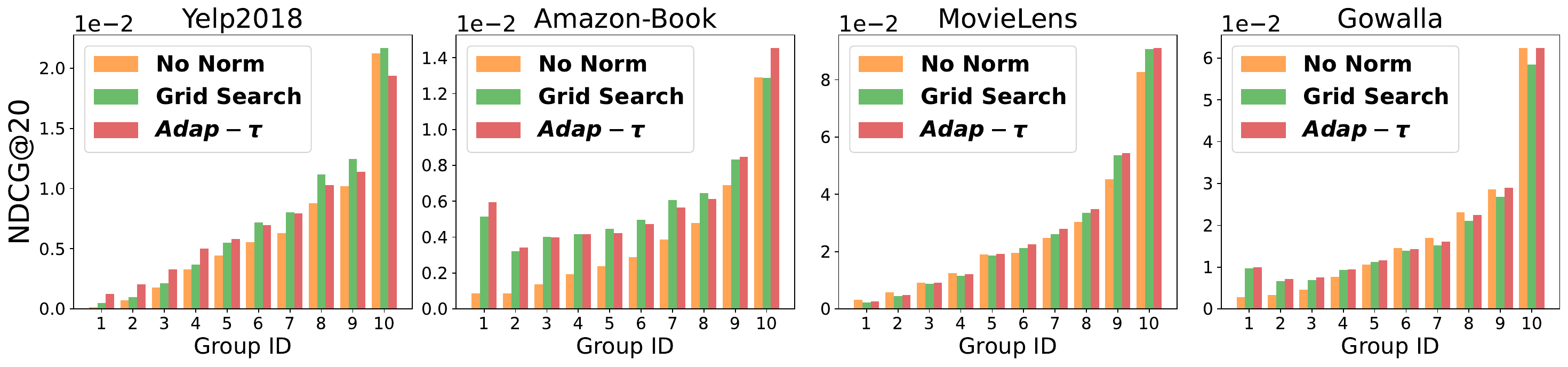}
    \caption{Performance comparison over different item groups among different strategies.
    }
    \label{fig:longtail_sum}
    \vspace{-.8em}
    \end{figure*}
    

We adopt four real-world datasets, Yelp2018\cite{he2020lightgcn}, Amazon-book \cite{he2020lightgcn}, Movielens \cite{yu2020graph} and Gowalla \cite{he2017neural_a}, to evaluate our model.  Detailed settings refers to \ref{appendix:param}.

\subsubsection{\textbf{Baselines}}
We validate the effectiveness of our method on two representative backbones: MF and LightGCN. Six strategies are tested in our experiments.
Compare to our Adap-$\tau$, we also adopt the following strategies:
\begin{itemize}[leftmargin=*]
    \item No norm: Adopting inner product where the user/item embeddings have not been normalized.
    \item Grid Search $\tau$: Normalizing the embeddings into a specific value (\ie $1/\sqrt(\tau)$), where $\tau$ is specified via fine-grained grid search (\ie step-size=0.02). 
    \item C-$\tau$: Following a similar topic in computer vision \cite{wang2017normface}, and leveraging a neural network to directly learn $\tau$. 
    \item Cu-$\tau$: A stronger baseline, where we improve the above C-$\tau$ and leverage the neural network to model personalized $\tau$.
    \item Adap-$\tau_0$: Leveraging an automatic $\tau_0$ to avoid grid search.
    \item Adap-$\tau$: Enhancing Adap-$\tau_0$ with personalized temperatures $\tau_u$. 
\end{itemize}
Meanwhile, we compare the model with various types of SOTA models, range from SGL\cite{wu2021self}, SimpleX\cite{mao2021simplex}, SimSGL\cite{yu2022graph}, NCL\cite{lin2022improving}.


\begin{figure}[h]\centering
    \vspace{-.4em}
    \includegraphics[width=1.0\linewidth]{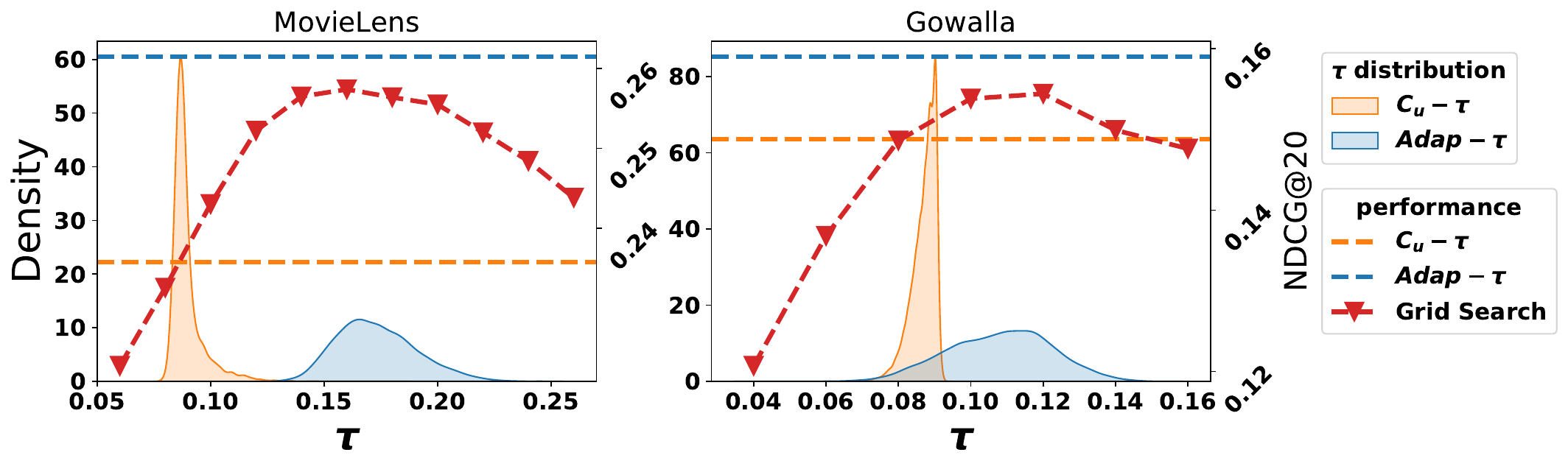}
    \caption{The dashed red curves denote the performance of the Grid Search $\tau$ with the different $\tau$. The dashed orange and blue lines indicate the performance of Adap-$\tau$ and Cu-$\tau$. We also report the distribution of the learned personalized $\tau$ for Adap-$\tau$ and Cu-$\tau$ as marked by the orange and blue regions.}
    \label{fig::tau_dist}
    \vspace{-1.em}
    \end{figure}
\subsubsection{\textbf{Parameter Settings}}
For a fair comparison, the embedding size is fixed to 64 and the initialization is unified with Xavier \cite{glorot2010understanding}. A grid search is conducted to confirm the optimal parameter setting for each model. Detailed implementation refers to \ref{appendix:param}.

\subsection{Performance Comparsion (RQ1)}
In this subsection, we conduct multiple experiments to validate the effectiveness of our strategy.
\subsubsection{\textbf{Effectiveness of our strategy}}
As can be seen from Tab. 3, with few exceptions, Adap-$\tau_0$ and Adap-$\tau$ that do not utilize any hyperparameter tuning on $\tau$, consistently outperforms the grid search baseline in all datasets and backbones. This result is highly encouraging, suggesting the limitation of embedding normalization can be obviated.

\subsubsection{\textbf{Impact of Adaptive Temperature}} 
From Table \ref{tab:main}, we observe that, Adap-$\tau$ obtains a superior performance against Cu-$\tau$ and C-$\tau$. We attribute this phenomenon to that C-$\tau$ is highly sensitive to initialized value and lack of benchmarked $\tau$ to decide proper distribution of $\tau$ (\cf Fig. \ref{fig::tau_dist}). Cu-$\tau$ and C-$\tau$ lack of critical supervisory signal to control the problem of gradient vanishment. Hence, it still exhibits inferior performance than our model.

\begin{figure*}[t]\centering
    \vspace{-.5em}
    \includegraphics[width=0.9\linewidth]{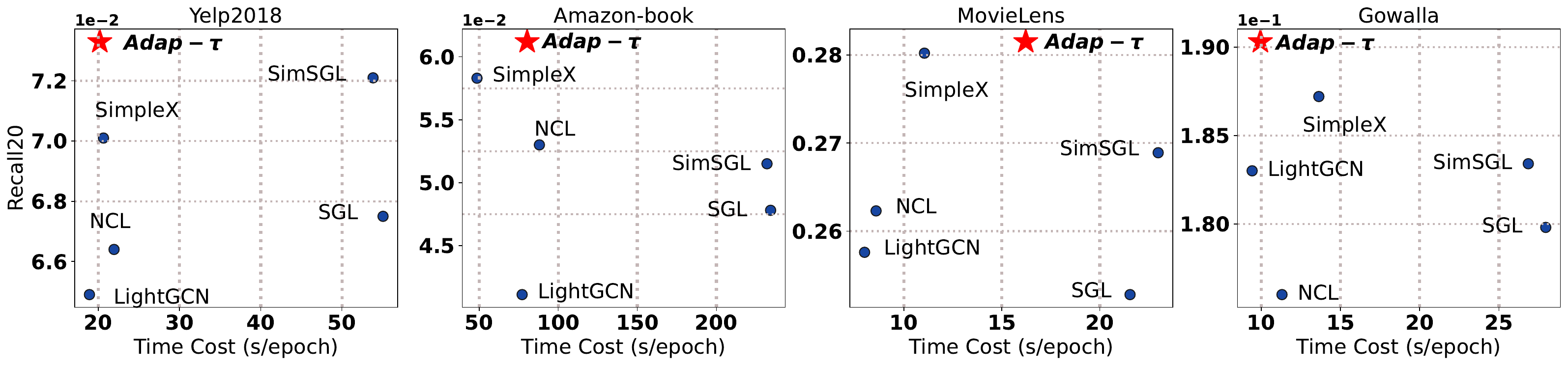}
    \caption{
    Performance comparisons in terms of both recommendation accuracy and efficiency.
    }
    \label{fig::sota_compare}
    \vspace{-.8em}
\end{figure*}
\subsubsection{\textbf{Impact of Fine-grained Temperature}}
By comparing with Adap-$\tau$ and Adap-$\tau_0$, we observe that Adap-$\tau$ consistently outperform Adap-$\tau_0$. This result indicates the superiority of leveraging personalized $\tau$ in RS. Once various users exhibit distinct rules or individual property, it is intuitively hard to control them by a fixed $\tau$. Similar results can be observed by Cu-$\tau$ outperforming C-$\tau$. 


\subsubsection{\textbf{Fair Recommendation.}}
As we mentioned in the introduction, we argue that the model without normalization will aggravate popularity bias. In this subsection, We give more detailed analyses from a fairness perspective. To prove that, we follow \cite{wu2021self}, splitting items into ten groups \textit{w.r.t.} their interaction frequency. adding normalization with Grid Search $\tau$ or adopting our Adap-$\tau$ can relieve the tensely contradictory relationship between long-tail and task of the normal recommendation. As we observe in Fig \ref{fig:longtail_sum}, the bar of "Grid Search $\tau$" and "Adap-$\tau$" surpass the counterpart by a considerable margin in smaller GoupID. It could also verify that our model has the capability of popularity debias from the side.

\begin{table}
    \small
    \caption{Results of MF under different ratio "noisy data"}
    \label{tab:robustness}
    \centering
    \begin{tabular}{cccccc}
        \toprule
        \multirow{2}{*}{ratio} & \multirow{2}{*}{model} & \multicolumn{2}{c}{Yelp2018} &  \multicolumn{2}{c}{Amazon-book} \\
        & &Recall  & NDCG & Recall  & NDCG  \\
        \hline
        \multirow{2}{*}{0.1} & Grid Search & 0.0722 & 0.0601 & 0.0564 & 0.0455  \\
        & Adap-$\tau$ & \textbf{0.0735} & \textbf{0.0613}  & \textbf{0.0577} & \textbf{0.0467} \\
        \hline
        \multirow{2}{*}{0.2} & Grid Search & 0.0703 & 0.0584 & 0.0534 & 0.0432  \\
        & Adap-$\tau$ & \textbf{0.0717} & \textbf{0.0593}  & \textbf{0.0546} & \textbf{0.0443} \\
        \hline
        \multirow{2}{*}{0.3} & Grid Search & 0.0696 & 0.0577 & 0.0509 & 0.0409  \\
        & Adap-$\tau$ & \textbf{0.0702} & \textbf{0.0584}  & \textbf{0.0520} & \textbf{0.0422} \\
        \hline
        \multirow{2}{*}{0.4} & Grid Search & 0.0678 & 0.0563 & 0.0493 & 0.0400 \\
        & Adap-$\tau$ & \textbf{0.0685} & \textbf{0.0569}  & \textbf{0.0507} & \textbf{0.0412} \\
        \hline
        \multirow{2}{*}{0.5} & Grid Search & 0.0667 & 0.0554 & 0.0481 & 0.0388 \\
        & Adap-$\tau$ & \textbf{0.0672} & \textbf{0.0560}  & \textbf{0.0487} & \textbf{0.0394} \\
       
        \bottomrule
    \end{tabular}
    \vspace{-1.8em}
\end{table}

\subsection{Adaptiveness Exploration (RQ2)}
Through previous experiments, we have realized our Adap-$\tau$ could adapt to the different datasets, backbone models, and users. In this section, we take a step further and explore how Adap-$\tau$ adapt to the data and users with different ratio of noise. Adding noises to the data would significantly change the data distribution, gradient magnitude, as well as the level of user hardness. We believe exploring the performance of Adap-$\tau$ on such a challenging task would help us further understand the adaptivity of Adap-$\tau$.

In this section, we exploit the Adaptiveness of our model over different "noisy data". Two strategies are adopted to add noise to the datasets. 1) In terms of the interaction frequency per user, we added false-positive items at the same proportion. 2) Splitting the users into four groups randomly, and we add fake items at a specific proportion according to the group ID. Strategy 1 concentrates on the \textbf{global adaptiveness} confronted with the same ratio of noisy data, while strategy 2 focuses on the \textbf{local adaptiveness} with respect to the "noisy ratio" individually.


\begin{figure}[t]\centering
\vspace{-.5em}
\includegraphics[width=.84\linewidth]{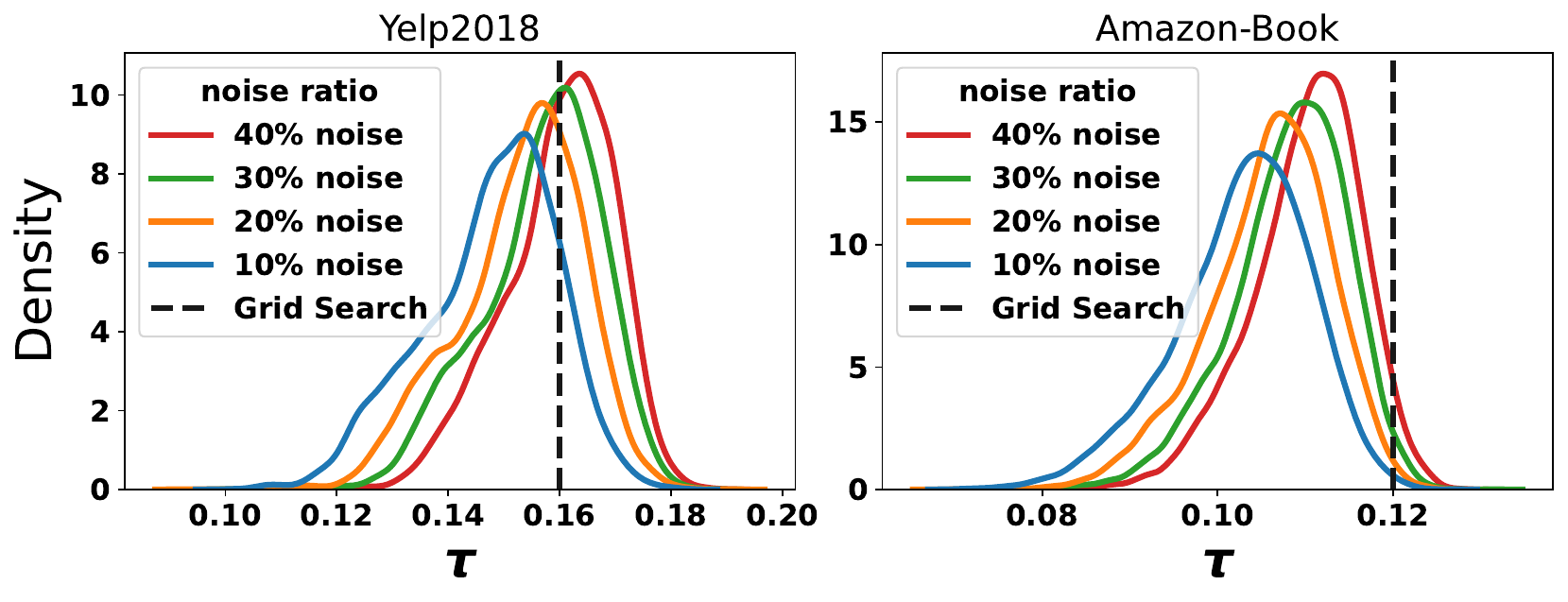}
\caption{
The distribution of $\tau$ with different ratio of noise data. Here noisy data is added via personalized manner, which different users are affected by different noise ratio.
}
\label{fig::dirty_2}
\vspace{-1.8em}
\end{figure}

\subsubsection{\textbf{Global Adaptiveness}}
When we focus on the result over strategy one, we observe that in Table \ref{tab:robustness}, the more noisy data added into training dataset, the larger $\tau$ will be chosen by grid searching. Meanwhile, our Adap-$\tau$ utilizes the feedback of each user and adaptively adjusts the $\tau$ to balance the hard-mining, which verifies its robustness against noise data and flexibility. Experiments also show its superior performance.

\subsubsection{\textbf{Local Adaptiveness}}
As we randomly split items into four groups and add 10\%,20\%,30\%,40\% ratio of noise data, the distribution of $\tau$ in each group shows the respective order in Fig \ref{fig::dirty_2}. Furthermore, with the rising amount of noise data, the adaptive $\tau$ also goes steadily up. And regardless of the ratio of noise data and the difference in adding strategy, our model achieves competitive results against hyperparameter searching without any tuning.

\subsection{Comparison with SOTA (RQ3)}
In this subsection, we are curious about how our model compares to those state-of-the-art models. Here, we choose two representative and powerful baseline: SGL\cite{wu2021self}, SimpleX\cite{mao2021simplex}, SimSGL\cite{yu2022graph}, NCL\cite{lin2022improving}, where SimpleX claims it surpass over 11 benchmark datasets and compared with 29 existing CF models in total.

From Figure \ref{fig::sota_compare}, we can see our proposed model Adap-$\tau$ obtains competitive results compared with state-of-the-art consistently. Meanwhile, we compare the time cost of each model, verifying that our method including the calculation of $\tau_0$ brings low time cost.

\section{Related work}
\subsection{Recommendation System}
The basic task of recommendation system is to predict potential interaction, which is called collaborative filtering. Existing methods could be roughly divided into three categories: MF-based methods \cite{hu2008collaborative,van2018representation}, VAE-based methods \cite{liang2018variational, ma2019learning, steck2019embarrassingly} and GNN-based methods \cite{he2020lightgcn, ying2018graph,wang2019neural}. GNN-based methods are inclined to achieve state-of-the-art performance with the development of Graph Neural Networks. For example, PinSage \cite{ying2018graph} borrows the idea from GraphSage while NGCF \cite{wang2019neural} devised NGCF. As the particularity of CF, LightGCN \cite{he2020lightgcn} throws away heavy and burdensome operations to show critical factors in the aggregation mechanism. However, the inner product represents the traditional measurement in the above models, while limited research tends to analyze its impact combine CF task. 

\subsection{Temperature and Normalization}
Temperature has exhibited its capability in numerous fields such as CV and NLP in particular with contrastive learning \cite{chen2020simple, he2020momentum}.
Motivated by the success in other areas, recommendation combined with contrastive learning has received scant attention in recent research literature \cite{mao2021simplex, tang2021multisample, wu2021self, zhou2021contrastive, wu2022effectiveness}. Although normalizaiton and $\tau$ are heuristically used by a small amount of work, it still lacks comprehensive exploration in recommendation. \cite{mao2021simplex} is mostly related work which analyzes the existing loss function and proposes a cosine contrastive loss to achieve hard-mining. However, its loss function filters easy items and needs detailed hyper parameter (margin and weight) searching, which restricted its flexibility in application and different datasets. To our best of knowledge, \cite{wu2022effectiveness} is the first work which utilize softmax loss directly into recommendation task. Despite the success of softmax has verified in recommendation system, how to understand it deeply and comprehensively with RS still remains challenge. As \cite{wu2022effectiveness} can not combine MF and softmax loss perfectly, here, our work focus on the softmax loss along with basic backbone from the pespective of theory and experiments.

\section{Conclusion and Future Work}
In this work, we focus on the embedding magnitude in recommendation system. With theoretical and empirical analysis, we emphasize the importance of embedding normalization. And we point out of the issue of straightforward normalization. Hence, we propose two principles to guide the adaptive learning of $\tau$. On these basis, we develop an adaptive and personalized $\tau$ without repeated searching over $\tau$ among different datasets. Experiments verify that our simple method is effective with different backbone in numerous dataset. 

Embedding magnitude is overlooked in many areas. We believe this study could draw researchers' attention on this issue and inspire more work on this line. It would be interesting to explore fine-grained temperature in other fields.

\begin{acks}
This work is supported by the National Key Research and Development Program of China (2021YFF0901603), the National Natural Science Foundation of China (62102382, U19A2079, 62121002), the CCCD Key Lab of Ministry of Culture and Tourism and the Starry Night Science Fund of Zhejiang University Shanghai Institute for Advanced Study (SN-ZJU-SIAS-001).
\end{acks}

\bibliographystyle{ACM-Reference-Format}
\bibliography{sample-base}

\appendix
\newpage
\section{Experimental Settings} \label{appendix:exp1}
\subsection{\textbf{Datasets and Metrics}}
\label{appendix:param}
\subsection{Parameter Settings.}
We implement our model in PyTorch and will release our implementation (codes, parameter setting, and training log) to enhance reproducibility. For a fair comparison, the embedding size is fixed to 64 for all methods and the initialization is unified with Xavier \cite{glorot2010understanding}. A grid search is conducted to confirm the optimal parameter setting for each model. To be more specific, learning rate is tuned among {1e-3, 5e-3, 1e-4} and the coefficient of $L_2$ regularization term is searched in $\{1e^{-9},1e^{-8},...,1e^{-1}\}$. As for the backbone of LightGCN, the number of layers is tuned among {1,2,3}, where dropout is adopted or not to prevent over-fitting. Focus on the traditional contrastive loss, temperature $\tau$ is a fine-grained search with an interval of 0.02. And the number of negative sampling is varying in {200, 400, 800, 1500}. Note that we perform SimpleX baseline in a detailed grid search accumulated more than 3500 experiments per dataset, where margin searched among $\{0.1,0.3,0.5,0.7,0.9\}$ and weight tuning among $\{50, 100,150,200,400,800\}$.
\subsection{Datasets and Metrics.}
We adopt four real-world datasets, Yelp2018\cite{he2020lightgcn}, Amazon-book \cite{he2020lightgcn}, Movielens \cite{yu2020graph} and Gowalla \cite{he2017neural_a}, to evaluate our model. For pair comparison, the Amazon-book, Yelp2018, and Gowalla are exactly the same as \cite{he2020lightgcn} used. The MovieLens is from \cite{yu2020graph} which is collected from the website \textit{movielens.umn\\.edu} and we use the version of 1M. Following \cite{he2020lightgcn, wang2019neural}, we leverage the routine strategy --- 10-core setting to preprocess the dataset. After standardization, we report the statistics of the above dataset in Tab. \ref{tab:datasets}. As for evaluation metrics, we adopt all-ranking protocol to compute recall@20 and ndcg@20.

\begin{table}[h]
    \centering
    \small
    \caption{Statistics of the datasets}
    \begin{tabular}{c|c|c|c|c}
    \toprule
        Dataset & \#Users & \#Items & \#Interactions & Density \\
        \hline
        \hline
         Yelp2018 & 31,831 & 40,841 & 1,666,869 & 0.0128\% \\
         Amazon-Book & 52,643 & 91,599 & 2,984,108 & 0.062\% \\
         Movielens & 6,022 & 3,043 & 995,154, & 5.431\% \\
         Gowalla & 29,858 & 40,981 & 1,027,370 &0.084\%\\
    \bottomrule
    \end{tabular}
    \vspace{-2.4em}
    \label{tab:datasets}
\end{table}
\section{Proofs}
\subsection{Proof of the lemma \ref{lam0}}
\label{proof_lama}
\begin{proof}

Let $\mathcal{L}(u)$ be the softmax loss on user $u$, and let notation $p_{ui}(\tau)$ be the logit of the instance $(u,i)$ governed by the parameter $\tau$, \ie
\begin{equation}
    \mathcal{L}(u)=-\sum_{i \in\mathcal{P}_u} {\frac{\exp{(f(u,i))}}{\sum_{j\in \mathcal I} \exp{( f(u,j))}}}
\end{equation}
\begin{align}
    p_{ui}(\tau)=\frac{exp(\frac{f(u,i)}{\tau})}{\sum\limits_{j\in \mathcal I}{\exp(\frac{f(u,j)}{\tau})}}
\end{align}
Note that the gradient  $\frac{{\partial L}}{{\partial f(u,i)}}$ can be written as:
\begin{align}
\label{gd1_}
    \frac{{\partial L}}{{\partial f(u,i)}} = \left\{ {\begin{array}{*{20}{c}}
        \frac{1}{\tau} {{p_{ui}(\tau)}(1 - \sum\limits_{k \in {N_u}} {{p_{uk}(\tau)}} )},  \quad \text{for} \quad y_{ui}=1\\
        - \frac{1}{\tau} {{p_{ui}(\tau)}(\sum\limits_{k \in {N_u}} {{p_{uk}(\tau)}} )}, \quad \text{for} \quad y_{ui}=0
        \end{array}} \right.
\end{align}
Owing to items $i$ are combined with two parts: $y_{ui}=0$ and $y_{ui}=1$, we shall represent it as a single expression for ease of subsequent lemma derivation.
\begin{align}
    \frac{{\partial L}}{{\partial f(u,i)}} = \frac{1}{\tau}\big(p_{ui}(\tau) \mathbb{I} [y_{ui}=1] - {p_{ui}(\tau)}(\sum\limits_{k \in {N_u}} {{p_{uk}(\tau)}} )\big)
\end{align}
where $\mathbb{I} (y_{ui}=1)$ represents an indicator function.
In terms of the normal gradient descent, we have 

\begin{equation}\label{gd}
\begin{aligned}
    \delta_{i} &= || e_i + \eta \frac{\partial L(u)}{\partial i} ||^2 - || e_i ||^2 \\
    &= || e_i + \eta \frac{\partial L(u)}{\partial f(u,i)} 
    \frac{\partial  f(u,i)}{\partial i}
    ||^2 - || e_i ||^2 \\
    &= 2 e_i^T (\eta \sum_{u} \frac{ \partial L(u)}{\partial i} e_u)  + o({\eta}^2\cdot \frac{\partial L(u)}{\partial i}) \\
    &=\sum_u \frac{2}{\tau} (\mathbb{I}[y_{ui}=1] - \sum\limits_{k \in {N_u}} {{p_{uk}(\tau)}} ) p_{ui} f(u,i) \\
\end{aligned}
\end{equation}
We can draw an observation from Lamma \ref{lam0}:
Note that at the early stage of the training procedure, users and items are distributed uniformly. In other words, $p_{ui} f(u,i)$ cannot tell remarkable difference and $\sum\limits_{k \in {N_u}} {{p_{uk}(\tau)}}$ is relatively small, while the magnitude of popular items will obtain explosive rising in term of $|\mathcal{P}_i| (\mathbb{I}[y_{ui}=1])$. That is:
\begin{equation} 
    \delta_{i} \propto |\mathcal{P}_i|
\end{equation}

Combine with above observation, we can prove the LEMMA \ref{lam0}.

\end{proof}
\subsection{Proof of Lemma \ref{la1}}
\label{proof_lemma2}
\begin{proof}
    Note that the gradient  $\frac{{\partial L}}{{\partial f(u,i)}}$ can be written as:
    \begin{align}
    \label{gd1}
        \frac{{\partial L}}{{\partial f(u,i)}} = \left\{ {\begin{array}{*{20}{c}}
           \frac{1}{\tau} {{p_{ui}(\tau)}(1 - \sum\limits_{k \in {N_u}} {{p_{uk}(\tau)}} )},  \quad \text{for} \quad y_{ui}=1\\
           - \frac{1}{\tau} {{p_{ui}(\tau)}(\sum\limits_{k \in {N_u}} {{p_{uk}(\tau)}} )}, \quad \text{for} \quad y_{ui}=0
            \end{array}} \right.
    \end{align}
    Thus, the expression (\ref{eq:t0}) can be transformed into:
    \begin{equation}
    \begin{aligned}
    \mathbb{E}_{u,i}[|\frac{{\partial L}}{{\partial f(u,i)}}|] 
        &=\frac{1}{m\tau}\mathbb{E}_{u}[2\sum\limits_{i\in  N_u}{p_{ui}(\tau)(1-\sum\limits_{k\in N_u}{p_{uk}(\tau)})}] \\
            &\leq \frac{2}{m\tau}(\mathbb E_u[\sum\limits_{i\in  N_u}{p_{ui}(\tau)}]-\mathbb E^2_u[\sum\limits_{i\in N_u}{p_{ui}(\tau)}]) \\
            &\leq \frac{2}{mt}(\mathbb E_u[\sum\limits_{i\in  N_u}{p_{ui}(\tau)}]-\mathbb E^2_u[\sum\limits_{i\in N_u}{p_{ui}(\tau)}]) \label{eq:qua}
        \end{aligned}
    \end{equation}
    where Cauchy Inequality is employed. The upper bound has a quadratic form. The optimal condition can be easily obtained by transforming the expression into $-(\mathbb E_u[\sum_{i\in  N_u}{p_{ui}(\tau)}]-\frac{1}{2})^2+1$.
    \end{proof}

\subsection{Proof of the lemma \ref{la2}}
\label{proof_lemma3}
\begin{proof}    
    For convenient, for each user $u$, let $a_u$ be the sum of rescaled prediction over his positive instances $\sum_{i\in N_u}{exp(f(u,i)/\tau)}$, and $b_u$ be the sum over all instances $\sum_{i}{{exp(f(u,i)/\tau)}}$. The expression of equation (\ref{eq:con}) can be transformed into:
    \begin{equation}
        \begin{aligned}
    \mathbb E_u[\sum\limits_{i\in N_u} {p_{ui}(\tau)}]
    &=\frac{\mathbb E_u[a_u]}{\mathbb E_u[b_u]}+\frac{\mathbb E_u[\frac{a_u}{b_u}] \mathbb E_up[b_u]-E[a_u]}{\mathbb E_u[b_u]} \\
    &=\frac{\mathbb E_u[a_u]}{\mathbb E_u[b_u]}-\frac{Cov(\frac{a_u}{b_u},b_u)}{E_u[b_u]}\\
    &\approx \frac{\mathbb E_u[a_u]}{\mathbb E_u[b_u]}
    \end{aligned}
    \end{equation}
    where $Cov(\frac{a_u}{b_u},b_u)$ denotes the covariance between the variables, which is bounded by the variance of $b_u$, \ie $Cov(\frac{a_u}{b_u},b_u)\leq \text{Var}_u[b_u]$. Considering in practice the value of $\text{Var}_u[b_u]$ is usually quite small, here we simply drop out the covariance term for derivation. 
    
    Now we turn to deal with the expression of $\mathbb E_u[a_u]$ and $\mathbb E_u[b_u]$.  Based on Taylor's expansion of an exponential function, we have:
    \begin{equation}
        \begin{aligned}
            {\mathbb E_u}[{b_u}] &= m{\mathbb E_f}[\exp (\frac{{\bf{f}}}{\tau })] \\
             &= m\exp ({\mathbb E_{\bf{f}}}[\frac{{\bf{f}}}{\tau }])(1 + \frac{{{\mathbb E_{\bf{f}}}[{{({\bf{f}} - \mu )}^2}]}}{{{\tau^2}2!}} + \sum\limits_{k = 3}^\infty  {\frac{{{\mathbb E_{\bf{f}}}[{{({\bf{f}} - \mu )}^k}]}}{{{\tau ^k}k!}}} )
        \end{aligned}
    \end{equation}
    when $\tau>2\lambda$, it can be approximated as:
    \begin{equation}
        \begin{aligned}
    E_u[b_u]\approx m \exp (\frac{\mu}{\tau} ) (1 + \frac{{{E_{\bf{f}}}[{{({\bf{f}} - \mu )}^2}]}}{{{\tau ^2}2!}}) \label{eq:apt}
    \end{aligned}
    \end{equation}
    since the higher-order term is bounded with:
    \begin{equation}
        \begin{aligned}
    |\sum\limits_{k = 3}^\infty  {\frac{{{E_{\bf{f}}}[{{({\bf{f}} - \mu )}^k}]}}{{{\tau ^k}k!}}| \le } \sum\limits_{k = 3}^\infty  {\frac{{2{{(\lambda /2)}^k}k!}}{{{\tau ^k}k!}}} = \frac{{2{{(\frac{\lambda }{{2\tau }})}^3}}}{{1 - \frac{\lambda }{{2\tau }}}} \le \frac{1}{{24}}
    \end{aligned}
    \end{equation} 
    where we use the fact that the central moment of a sub-exponential variable is bounded:
    \begin{equation}
        \begin{aligned}
        {E_{\bf{f}}}[|({\bf{f}} - \mu ){|^k}] &= \int_0^\infty  {p({\rm{|}}({\bf{f}} - \mu ){|^k} > t)dt} \\
         &= \int_0^\infty  p \left( {{\rm{|}}({\bf{f}} - \mu ){|^k} > {t^{1/k}}} \right){\rm{d}}t\\
         &\le \int_0^\infty  2 {e^{ - \frac{{2{t^{1/k}}}}{\lambda }}}{\rm{d}}t \\
         &= 2{(\lambda /2)^k}k\int_0^\infty  {{e^{ - u}}} {u^{k - 1}}\;{\rm{d}}u = 2{(\lambda /2)^k}k!
    \end{aligned}
    \end{equation} 
    Equation (\ref{eq:apt}) can be further transformed into:
    \begin{equation}
        \begin{aligned}
    {E_u}[{b_u}] \approx m\exp (\frac{\mu }{\tau })\exp (\frac{{{E_{\bf{f}}}[{{({\bf{f}} - \mu )}^2}]}}{{2{\tau ^2}}})
    \end{aligned}
    \end{equation} 
    as $\exp (\frac{{{E_{\bf{f}}}[{{({\bf{f}} - \mu )}^2}]}}{{2{\tau ^2}}}) = 1 + \frac{{{\sigma ^2}}}{{2{\tau ^2}}} + o({(\frac{{{\sigma ^2}}}{{2{\tau ^2}}})^2})$ and $\frac{{{\sigma ^2}}}{{2{\tau ^2}}}\leq \frac{{{\lambda ^2}}}{{2{\tau ^2}}} \le \frac{1}{8}$.
    Similar treatment can be conducted for $E_u[a_u]$ and we can get:
    \begin{equation}
        \begin{aligned}
    {E_u}[{a_u}] \approx \frac{|D|}{n}\exp (\frac{\mu_+ }{\tau })\exp (\frac{\sigma_+^2}{{2{\tau ^2}}})
    \end{aligned}
    \end{equation} 
    
    Thus, the original conditional equation can be transformed into:
    \begin{equation}
        \begin{aligned}
            \frac{{{E_u}[{a_u}]}}{{{E_u}[{b_u}]}} \approx \frac{{\frac{{|D|}}{n}\exp (\frac{{{\mu _ + }}}{\tau })\exp (\frac{{\sigma _ + ^2}}{{2{\tau ^2}}})}}{{m\exp (\frac{\mu }{\tau })\exp (\frac{{{\sigma ^2}}}{{2{\tau ^2}}})}} = \frac{1}{2}
    \end{aligned}
    \end{equation} 
    With the reorganization, we can find the equation has a quadratic form \textit{w.r.t.} $1/\tau$ and thus can write the root of the equation as:
    \begin{equation}
        \begin{aligned}
    \tau\approx \frac{{\sigma _ + ^2 - {\sigma ^2}}}{{ - ({\mu ^ + } - \mu ) + \sqrt {{{({\mu ^ + } - \mu )}^2} + 2(\sigma _ + ^2 - {\sigma ^2})\log (\frac{{nm}}{{2|D|}})} }}
    \end{aligned}
    \end{equation} 
    When $\sigma _ + ^2$ is quite close to $\sigma^2$, $\tau_0$ can be approximated with:
    \begin{equation}
        \begin{aligned}
            \tau_0\approx\frac{\mu_{+}-\mu}{\log(\frac{nm}{2|\Set D|})} 
            \end{aligned}
        \end{equation}
    The lemma gets proofed.
\end{proof}
\section{Validation on approximation}
\subsection{Approximation \wrt $\tau_0$}
In our implementation, we pursue efficiency so as to simplify the expression of $\tau_0$. To verify its justifiability, we record the real value of  $\sigma_+^2 - \sigma^2$  and corresponding $\tau_0$. Actually, we could obtain similar performance under Eq. \ref{eq:sol1} and Eq. \ref{eq:sol}.
\label{proof_apendxi}
\begin{table}[h]
    \caption{Comparison between approximation on $\tau_0$.}
    \label{tab:proof_}
    \centering
    \begin{tabular}{c|c|c|c}
        \toprule
        Datasets & $\sigma_+^2 - \sigma^2$ & $\tau_0$ by Eq. \ref{eq:sol1}&  $\tau_0$ by Eq. \ref{eq:sol}\\
        \hline
         Yelp2018 & -0.004362 & 0.095602 &  0.099263 \\
         Amazon-Book & -0.001356 & 0.078836 &  0.079994 \\
         MovieLens & -0.017910 & 0.148159 &  0.156905 \\
         Gowalla & -0.001148 & 0.094281 &  0.095187 \\
        \bottomrule
    \end{tabular}
\end{table}
\subsection{Gradient Vanishment}
\begin{figure}[h]\centering
    \vspace{-.5em}
    \includegraphics[width=1.0\linewidth]{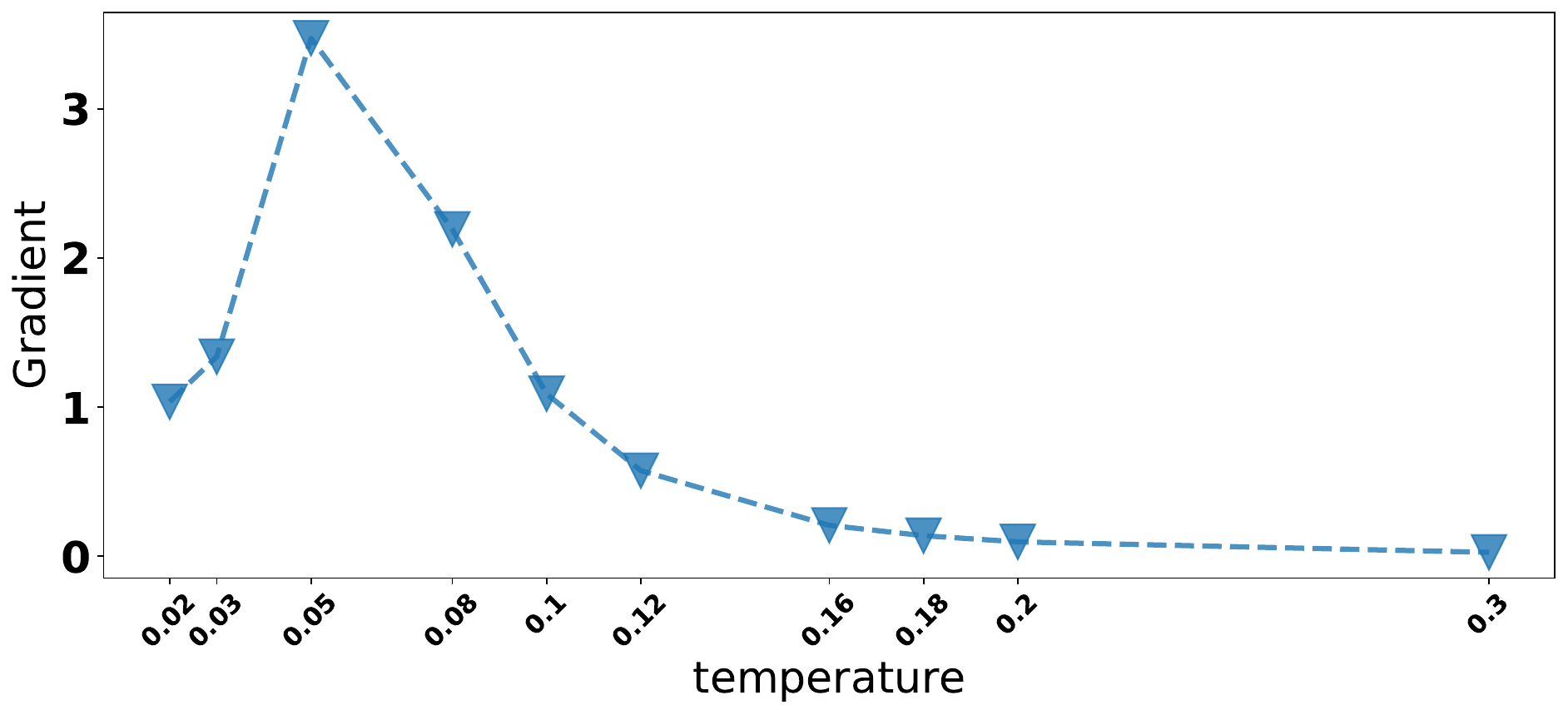}
    \caption{
    The gradient \wrt $\tau$ on Gowalla.
    }
    \label{vannise_proof}
    \vspace{-1.8em}
    \end{figure}

To verify the impact of $\tau$ on the gradient vanishment, we adopt different temperatures to train the model and record the respective gradient according to Eq. \eqref{eq:vanish222}. Figure \ref{vannise_proof} verifies its effectiveness on gradient. In particular, when $\tau$ is larger than $0.3$, the overall gradient is nearly zero. And the peak of Figure \ref{vannise_proof} is equivalent to the best $\tau$ via grid search.

\label{sec_vannise_proof}
\newpage
\section{PSEUDO-CODE FOR Adap-$\tau$}
\label{algorithm_pse}
\begin{algorithm}[h]
    \caption{\label{alg:main} Adap-$\tau$'s main learning algorithm.}
    \begin{algorithmic}
        \State \textbf{input:} Number of users and items:n,m; Users and Items in training set $U^{\prime},I^{\prime} \in \mathcal{D}$; Batch size: B; Number of neagative sampling: M;Dimension: d.
        \For{epoch $\in \{1, \ldots, N\}$}
        \State \textcolor{gray}{~~~~~~~~~~~~~~~~~~~~~~~~~~~~~~~~~~~~~~~~\# 1. Compute $\tau_0$}
        \State $~~~~$ Get user and item embeddings $e(U^{\prime}), e(I^{\prime})$
        \State $~~~~$ $e(U^{\prime}), e(I^{\prime})=normalize(e(U^{\prime})), normalize(e(I^{\prime}))$
        \State \textcolor{gray}{~~~~~~~~~~~~~~~~~~~~~~~~~~~~~~~~~~~~~~~~\# Dimension $e(U^{\prime})$: [n, d], $e(I^{\prime})$: [m, d]}
        \State $~~~~$ $\mu_{+}= mean(\bm e(U^{\prime})^\top \bm e(I^{\prime}))$ \Comment{\textcolor{gray}{~~~~~~~~~~~~~~~~~~~~~~~~~~~~~~ $O(|\mathcal{D}|d)$}}
        \State $~~~~$ $\mu=   mean(e(U^{\prime})^\top  mean(e(I^{\prime},dim=1)) )$ \Comment{\textcolor{gray}{~~~~~~~~~~~~~~~~~~~~~~~~~~~~~~ $O(nd)$}}
        \State $~~~~$ $\tau_0=\frac{\mu_{+}-\mu}{\log(\frac{nm}{2|\Set D|})} $ \Comment{\textcolor{gray}{~~~~~~~~~~~~~~~~~~~~~~~~~~~~~~ $O(1)$}}

        \State $~~~~$ Calculate acumulataed loss of each user $\hat{\mathcal{L}}(u)$ 
        \State \textcolor{gray}{~~~~~~~~~~~~~~~~~~~~~~~~~~~~~~~~~~~~~~~~\# 2. Start Training.}
        \For{sampled minibatch $\{\bm u,\bm i, \bm j\} $ in DataLoader}
        \State $~~~~$ Get user and item embeddins $e(u), e(i), e(j)$
        \State $~~~~$ $s({pos}) = \bm e_u^\top \bm e_i / (\lVert\bm e_u\rVert \lVert\bm e_i\rVert)$ \Comment{\textcolor{gray}{~~~~~~~~~~~~~~~~~~~~~~~~~~~~~~ postive score $O(Bd)$}}
        \State $~~~~$ $s(neg) = \bm e_u^\top \bm e_j / (\lVert\bm e_u\rVert \lVert\bm e_j\rVert)$\Comment{\textcolor{gray}{~~~~~~~~~~~~~~~~~~~~~~~~~~~~~~ negative score $O(BdM)$}}
        \State $~~~~$ Calculate $\tau_u$ according to $\hat{\mathcal{L}}(u), \tau_0$ as Eq. \eqref{eq:lw}
        \State \textbf{define} $\mathcal{L}(u)$ \textbf{as}~ \\ $\mathcal{L}(u) \!=\! -\log \frac{\exp(s_{u,i}(pos)/\tau_u)}{\exp(s_{u,i}(pos)/\tau_u) + \sum \exp(s_{u,j}(neg)/\tau_u)}$  \State $\mathcal{L} = \frac{1}{B} \sum_{u=1}^B \left[\mathcal{L}(u)\right]$
        \State update networks $f$ to minimize $\mathcal{L}$
        \EndFor
        \EndFor
        \State \textbf{return} encoder network $f(\cdot)$.
    \end{algorithmic}
\end{algorithm}
\end{document}